\documentclass[aps,pra,superscriptaddress,twocolumn]{revtex4-1}

\usepackage{graphicx,graphics,epsfig}   
\usepackage{dcolumn}    
\usepackage{bm}         
\usepackage{amsmath}    
\usepackage{verbatim}   
\usepackage{color}      
\usepackage{subfigure}  
\usepackage{amsmath,amsfonts,amssymb,amsthm,graphics,graphics,color,bbm}
\usepackage[rgb]{xcolor}
\usepackage{enumerate}
\usepackage{times}

\newcommand{\tr}{\mathrm{tr}}
\newcommand{\ket}[1]{|#1\rangle}
\newcommand{\bra}[1]{\langle#1|}

\newcommand{\ketbra}[2]{|#1\rangle\langle#2|}

\newcommand{\rA}{\mathrm{A}}
\newcommand{\rB}{\mathrm{B}}

\newcommand{\be}{\begin{eqnarray}}
\newcommand{\ee}{\end{eqnarray}}

\newtheorem{result}{Result}

\begin{document}

\title{Classical communication cost of quantum steering}

\author{Ana Bel\'en Sainz}\affiliation{H. H. Wills Physics Laboratory, University of Bristol$\text{,}$ Tyndall Avenue, Bristol, BS8 1TL, United Kingdom}
\author{Leandro Aolita}
\affiliation{Instituto de F\'isica, Universidade Federal do Rio de Janeiro, P. O. Box 68528, Rio de Janeiro, RJ 21941-972, Brazil}
\author{Nicolas Brunner}\affiliation{D\'epartement de Physique Th\'eorique, Universit\'e de Gen\`eve, 1211 Gen\`eve, Switzerland}
\author{Rodrigo Gallego}\affiliation{Dahlem Center for Complex Quantum Systems, Freie Universit\"at Berlin, 14195 Berlin, Germany}
\author{Paul Skrzypczyk}\affiliation{H. H. Wills Physics Laboratory, University of Bristol$\text{,}$ Tyndall Avenue, Bristol, BS8 1TL, United Kingdom}

\begin{abstract}
Quantum steering is observed when performing appropriate local measurements on an entangled state. Here we discuss the possibility of simulating classically this effect, using classical communication instead of entanglement. We show that infinite communication is necessary for exactly simulating steering for any pure entangled state, as well as for a class of mixed entangled states. Moreover, we discuss the communication cost of steering for general entangled states, as well as approximate simulation. Our findings reveal striking differences between Bell nonlocality and steering, and provide a natural way of measuring the strength of the latter.
\end{abstract}

\maketitle



The concept of steering, first introduced by Schr\"odinger \cite{schrodinger}, was recently put on firm grounds in a quantum information-theoretic setting \cite{wiseman07}. It describes the following genuinely quantum effect: consider an experiment with two distant observers, Alice and Bob, sharing an entangled quantum state. By performing a local measurement on her subsystem, Alice can remotely steer the quantum state of the system held by Bob. Quantum steering thus elegantly captures the celebrated ``spooky action at a distance'' discovered by Einstein-Podolsky-Rosen, and represents a fundamental form of nonlocality in quantum theory, intermediate between entanglement and Bell nonlocality \cite{wiseman07,quintino15}. It can be detected via the violation of so-called steering inequalities \cite{cavalcanti09} (analogous to Bell inequalities), and experimental demonstrations have been reported \cite{wittman}. Moreover, steering is directly connected to measurement incompatibility \cite{quintino14,uola14}, and offers applications in quantum information theory \cite{cyril,piani}. 

In order to gain insight into this strikingly counter-intuitive aspect of quantum theory, it is relevant to discuss quantum steering from the perspective of more general nonlocal resources. While the observation of steering (e.g. via violation of a steering inequality) rules out any explanation based on purely classical correlations (local resources), one may ask whether the use of an additional nonlocal resource, such as classical communication, could explain the phenomenon. In particular, it is then relevant to ask how much classical communication would be required in order to perfectly reproduce quantum steering.

The goal of the present is precisely to explore these questions. Specifically, we discuss the \emph{communication cost of quantum steering}, i.e., the minimal amount of classical communication required for reproducing the statistics of a given quantum steering experiment, without using any entanglement. The communication cost will generally depend on which entangled state is considered, and which measurements are performed. We believe that this provides a natural way of measuring the strength of quantum steering, complementary to previously introduced measures \cite{piani,skrzypczyk13,rodrigo14}. Moreover, this measure allows for a direct comparison of steering with other notions of nonlocality in quantum theory, in particular with Bell nonlocality \cite{bell64,review}. Indeed, the communication cost of simulating Bell nonlocal quantum correlations has received much attention \cite{maudlin,brassard99,steiner,gisingisin,tonerbacon,degorre,regev}. 

Here we show that the communication cost of perfectly simulating the steering of any pure entangled state, considering all possible projective measurements, is infinite. Interestingly, this is in strong contrast with the case of Bell nonlocality where few bits of communication suffice in several cases. For instance, the statistics of projective measurements on a two-qubit maximally entangled state can be simulated using only one bit of communication \cite{tonerbacon}. Then, we develop a general method for lower-bounding the communication cost of an arbitrary quantum steering experiment. Finally we discuss the communication cost of approximately simulating steering experiments, and conclude with some open questions.

\emph{Scenario.}--- Consider a game with two distant players, Alice and Bob, sharing a quantum state $ \varrho_{\rA\rB}$. The players want to convince a referee that the state they share is entangled (see Fig.1). In order to do so, the referee will perform the following test. He will ask Alice to perform a measurement $x$ on her half of the state, and announce its result $a$, while Bob is asked to send his subsystem to the referee. Then, by performing tomography over many instances of the game, the referee can characterize the conditional (unnormalised) state of Bob's subsystem, which is given by
\begin{align}\label{assemblage}
\sigma_{a|x} &= \tr_\rA\left[ \varrho_{\rA\rB} \, (M_{a|x} \otimes  \openone) \right], 
\end{align}
where $M_{a|x}$ denotes the POVM element (effect operator) of Alice corresponding to the outcome $a$ of the measurement setting $x$. The collection $\sigma=\{\sigma_{a|x}\}$ of conditional states $\sigma_{a|x}=p(a|x)\,\varrho_{a|x}$ is termed an assemblage; here $p(a|x)=\tr[\sigma_{a|x}]$ and $\varrho_{a|x}$ is a normalised quantum state. Importantly, one should enforce that Alice and Bob cannot communicate during the game (e.g. by ensuring space-like separation). 

Recently, Wiseman et al. \cite{wiseman07} discussed the above game in an information-theoretic setting, and showed formally how the referee can ensure that the players do indeed share entanglement. Specifically, they characterised the most general cheating strategy, referred to as a local hidden state (LHS) model. Consider the case in which Alice and Bob do not share an entangled state, but only classical correlations (or equivalently a separable quantum state) represented by a shared classical variable $\lambda$. Upon receiving measurement setting $x$ from the referee, Alice provides an outcome $a$ according to a response function $p(a|x,\lambda)$. At the same time, Bob sends a (normalised) quantum state $\varrho_{\lambda}$ (unentangled to Alice) to the referee. With such a strategy, the assemblage prepared for the referee by Alice and Bob necessarily has the form
\be\label{eq:lhs}
\sigma_{a|x} = \int d\lambda\,  \mu(\lambda)\, p(a|x,\lambda)\,  \varrho_{\lambda} ,
\ee
where $\mu(\lambda)$ represents the probability density of the shared variable $\lambda$, i.e. $\int d\lambda\,  \mu(\lambda)=1$. Assemblages of the above form are said to admit a LHS model. Therefore, if the assemblage held by the referee has a LHS model, he will not be convinced that Alice and Bob do indeed share entanglement, and Alice and Bob thus loose the game. On the contrary, if the referee can certify that the assemblage cannot be decomposed as above, he concludes that Alice and Bob share entanglement, and thus they win the game.

Interestingly, while entanglement is clearly necessary in order to demonstrate steering (i.e. win the game), it is in general not sufficient. That is, there exist entangled states for which Alice and Bob can never win the game, as the state admits a LHS model. For any possible measurement performed by Alice, the corresponding assemblage can be perfectly reproduced using only shared classical variables. This was first demonstrated when the referee only asks Alice to perform projective measurements, and recently extended to general POVMs. More generally, this line of research aims at understanding the relation between various forms of quantum correlations. 

In the present work, we consider LHS models augmented by classical communication between Alice and Bob. In other words, we consider more general cheating strategies. This scenario is motivated in various ways. First, classical communication represents the most natural classical resource for obtaining nonlocality. Hence, the minimal amount of classical communication required to reproduce a quantum assemblage provides a natural measure of the strength of steering. Moreover, as this approach has a long history of studies in the context of Bell nonlocality, it offers a natural way of quantitatively comparing quantum steering and Bell nonlocality. 

\begin{figure}
\includegraphics[width=0.65\columnwidth]{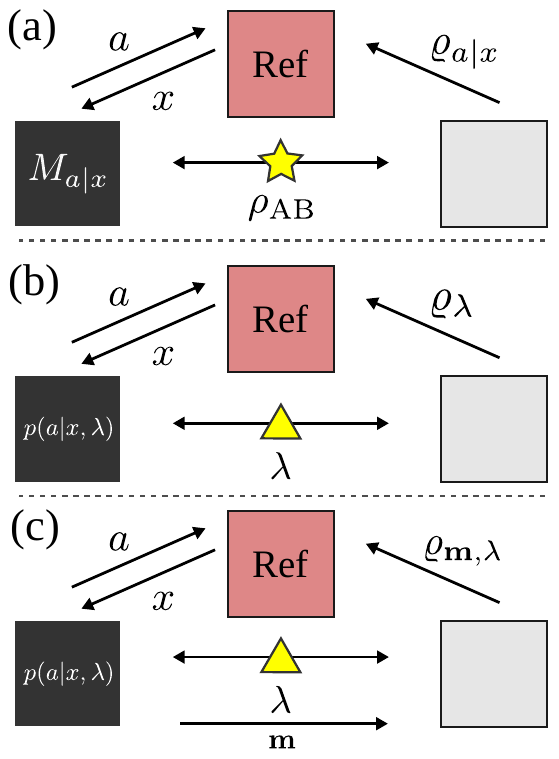}
\caption{\label{setup}  (a) Quantum steering experiment, (b) local hidden state strategy, (c) local hidden state strategy augmented with classical communication, the scenario investigated in the present work.}
\end{figure}

More formally, we consider the following cheating strategies. First, note that only communication from Alice to Bob is relevant, as the one from Bob to Alice is a free operation for steering, therefore useless to fake steering using unsteerable assemblages \cite{rodrigo14}. So, upon receiving her measurement setting $x$ from the referee, Alice is now allowed to send a classical message ${\bf m}$ of $t$ bits to Bob. This allows Bob to send a (normalised) quantum state $\varrho_{{\bf m},\lambda}$ to the referee, having now partial (or complete) knowledge about $x$. 
\be
\label{assem cc}
\sigma^{(\mathrm{sim})}_{a|x} =  \sum_{\bf m}  \int d\lambda \, \mu(\lambda)\, q({\bf m}|x,\lambda)\, p(a|x,\lambda)\,  \varrho_{{\bf m},\lambda} ,
\ee
where $q({\bf m}|x,\lambda)$ is the probability for Alice to send ${\bf m}$ given setting $x$ and shared variable $\lambda$. Clearly, LHS assemblages represent a subclass of the above ones. 


Finally, it is worth emphasizing that the cheating strategies we consider reproduce both the conditional quantum states $\varrho_{a|x}$ of Bob and the local statistics $p(a|x)$ of Alice.



\emph{Quantum states with infinite communication cost.}--- We first show the communication cost of steering is infinite for any pure entangled bipartite state. More precisely, we show the following:

\begin{result}
\label{theo1}
Consider that Alice and Bob share an arbitrary bipartite pure entangled state (of any dimension) and Alice performs local projective (rank-1) measurements. 
The resulting assemblage cannot be simulated using a LHS model augmented with finite communication; the length $t$ of the transmitted message ${\bf m}$ must be infinite, i.e. $t = \infty$.
\end{result}
\begin{proof}[Proof]
For clarity, we will detail the proof for the case of a two-qubit entangled state, $\ket{\psi_{\theta}} = \cos \theta\, \ket{00} + \sin \theta\, \ket{11}$, with $0 < \theta \leq \pi/4$. We consider arbitrary projective qubit measurements on Alice's side, hence we identify the measurement label $x$ with unit Bloch vectors $\hat{\mathbf{x}}$. The corresponding projectors are 
$\{\Pi_{0|\hat{\mathbf{x}}}, \Pi_{1|\hat{\mathbf{x}}}\}$, with $\Pi_{a| \hat{\mathbf{x}}}=\ket{\phi_{a|\hat{\mathbf{x}}}}\bra{\phi_{a|\hat{\mathbf{x}}}}$. Using Eq. \eqref{assemblage} we obtain that 
\begin{eqnarray}
\nonumber\sigma_{a|\hat{\mathbf{x}}}&=&\tr_A ( \ketbra{\phi_{a|\hat{\mathbf{x}}}}{\phi_{a|\hat{\mathbf{x}}}}\otimes \openone \ketbra{\psi_\theta}{\psi_\theta})\\
\label{eq:rank1}&=& p(a|\hat{\mathbf{x}})\ketbra{\Phi_{a|\hat{\mathbf{x}}}}{\Phi_{a|\hat{\mathbf{x}}}}
\end{eqnarray}
where $p(a|\hat{\mathbf{x}})=\tr ( \ketbra{\phi_{a|\hat{\mathbf{x}}}}{\phi_{a|\hat{\mathbf{x}}}}\otimes \openone \ketbra{\psi_\theta}{\psi_\theta})$, and $\ket{\Phi_{a|\hat{\mathbf{x}}}}= \bra{\phi_{a|\hat{\mathbf{x}}}} \otimes \openone \ket{\psi_\theta}/\sqrt{p(a|\hat{\mathbf{x}})}$ is the steered state. 
First, note that Eq. \eqref{eq:rank1} shows that $\sigma_{a|\hat{\mathbf{x}}}$ is proportional to a rank-1 projector. Second, the measurement setting runs over all possible Bloch vectors, hence it follows that the steered states $\ket{\Phi_{a|\hat{\mathbf{x}}}}$ run over all the possible pure qubit states.

Now, the most general assemblage that can be generated using an LHS model augmented with classical communication takes the form \eqref{assem cc}. Then, since the target assemblage consists of pure states and $\mu(\lambda)\,q({\bf m}|\hat{\mathbf{x}},\lambda)\geq0$ for all $\mathbf{m},\lambda$, each $\varrho_{\mathbf{m},\lambda}$ must be the same rank-1 projector, i.e. for each $a$ and $\hat{\mathbf{x}}$,
\begin{equation}
\label{rank-1_proj}
\varrho_{\mathbf{m},\lambda} = \ket{\Phi_{a|\hat{\mathbf{x}}}}\bra{\Phi_{a|\hat{\mathbf{x}}}}\, \forall\, \mathbf{m},\lambda \text{ s.t. } \mu(\lambda)\,q({\bf m}|\hat{\mathbf{x}},\lambda)>0.
\end{equation}
Recall furthermore that $\lambda$ is independent of $\hat{\mathbf{x}}$, having been distributed before Alice learns $\hat{\mathbf{x}}$ from the referee, and consider a particular value $\lambda^*$ which occurs with non-zero probability. If a message $\mathbf{m}$ of any finite length $t$ is sent from Alice to Bob, then there are at most $2^t$ distinct pure states $\varrho_{\mathbf{m},\lambda^*}$ (for each $\lambda^*$) that Bob can send to the referee. However, we already noted that the states $\ket{\Phi_{a|\hat{\mathbf{x}}}}\bra{\Phi_{a|\hat{\mathbf{x}}}}$ in the target assemblage run over infinitely many pure states. Therefore, no finite 
length message can reproduce the target assemblage. 
The extension to higher dimension uses the same argument (although note that the inputs are not described by Bloch vectors any more) and is straightforward.
 \end{proof}

A natural question following from the above, is to ask whether the communication cost of steering for more general entangled states, not only pure entangled states, could be infinite. Our second result is to show that indeed there exist mixed entangled states that also have infinite communication cost.

\begin{result}
\label{theo2}
Consider that Alice and Bob's shared state is the (normalised) projector onto the anti-symmetric subspace of two qudits, $\varrho_d^\mathrm{anti} = 2 A_d/d(d-1)$, where $A_d = \tfrac{1}{2}(\openone - \sum_{ij}\ket{ij}\bra{ji})$, and that Alice performs local projective (rank-1) measurements. 
The resulting assemblage cannot be simulated using a LHS model augmented with finite communication; the length $t$ of the transmitted message ${\bf m}$ must be infinite, i.e. $t = \infty$.
\end{result}
\begin{proof}[Proof]
Let us denote the projectors of the (arbitrary) measurement of Alice  $\Pi_{a|\hat{\mathbf{x}}}=\ket{\phi_{a|\hat{\mathbf{x}}}}\bra{\phi_{a|\hat{\mathbf{x}}}}$. Using Eq. \eqref{assemblage} with $\varrho_d^\mathrm{anti}$ we obtain 
\begin{eqnarray}
\sigma_{a|\hat{\mathbf{x}}}&=&\tfrac{1}{d(d-1)}(\openone-\ketbra{\phi_{a|\hat{\mathbf{x}}}}{\phi_{a|\hat{\mathbf{x}}}})
\end{eqnarray}
which has the property that each steered state of Bob is orthogonal to the measurement direction $\ket{\phi_{a|\hat{\mathbf{x}}}}$ of Alice. Since the measurement settings $\hat{\mathbf{x}}$ of Alice run over all projective measurements (i.e. bases), she can ensure that Bob's state is orthogonal to one of the states in any given basis. 

Again, the most general assemblage that can be generated using an LHS model augmented with classical communication takes the form \eqref{assem cc}. It follows that each $\varrho_{\mathbf{m},\lambda}$ must be orthogonal to the same rank-1 projector, i.e. for each $a$ and $\hat{\mathbf{x}}$,
\begin{equation}
\label{same_rank1_proj}
\bra{\phi_{a|\hat{\mathbf{x}}}}\varrho_{\mathbf{m},\lambda}\ket{\phi_{a|\hat{\mathbf{x}}}} = 0\, \forall\, \mathbf{m},\lambda \text{ s.t. } \mu(\lambda)\,q({\bf m}|\hat{\mathbf{x}},\lambda)>0.
\end{equation}
Identically to before, $\lambda$ is independent of $\hat{\mathbf{x}}$. If a message $\mathbf{m}$ of any finite length $t$ is sent from Alice to Bob, then there are at most $2^t$ distinct pure states $\varrho_{\mathbf{m},\lambda}$ (for each $\lambda$) that Bob can send to the referee. However, we already noted that the states $\ket{\phi_{a|\hat{\mathbf{x}}}}$ in the target assemblage to which Bob's simulated states need to be orthogonal run over infinitely many states. Therefore, since no finite set of states can be orthogonal to all states, no finite length message can reproduce the target assemblage.
 \end{proof} 

As a side remark complementing above results, note that infinite communication is always sufficient to simulate any assemblage with a straightforward strategy. Alice outputs $a$ according to $p(a|\hat{\mathbf{x}})$. She then sends the classical message $m = (\hat{\mathbf{x}},a$) to Bob, who then prepares and forwards $\varrho_{a|\hat{\mathbf{x}}}$ to the referee.

Notably, Results \ref{theo1} and \ref{theo2} show that the communication costs of quantum steering and of quantum nonlocality are totally different. In particular, while the communication cost of steering is infinite for any pure entangled two-qubit state, few bits of communication of enough in the context of nonlocality. Specifically, the statistics of local projective measurements on a maximally entangled state can be reproduced with a single bit of communication \cite{tonerbacon}, while two bits are enough for partially entangled states \cite{tonerbacon}. For higher dimensional states, it was shown that two bits of communication suffice to reproduce the correlations of dichotomic measurements on any bipartite entangled states \cite{regev}. Nevertheless, it is known that the statistics of general measurements on $d \times d$ maximally entangled states, require an amount of communication that increases (at least) as $O(d)$ \cite{brassard99}. 

It is also worth making a connection to entanglement theory. The certification of entanglement can be recast in the setting of Fig. \ref{setup}, by demanding now that Alice also sends a quantum state to the Referee (similarly to Bob). In this case, even infinite communication will not help the players, since entanglement cannot be created by LOCC, i.e. local operations assisted by an arbitrary amount of (possibly two-way) classical communication. An analogy can be drawn to semi-quantum games \cite{buscemi}Ê (where players receive quantum inputs) for which it was also demonstrated that infinite communication can never replace an entangled state \cite{Rosset13}. 

These results further confirm that steering can be considered a form of quantum nonseparability which is intermediate between entanglement and Bell nonlocality.

\emph{Communication cost for arbitrary assemblages.}--- A natural question following the above, is to discuss the communication cost of steering for more general entangled states. In particular, one may expect such cost to be finite for entangled states which are of full rank. To discuss this point, we provide a method for placing a lower bound on the communication cost for simulating an arbitrary assemblage, considering an arbitrary (but finite) number of measurements. 

\begin{result}
\label{theo3}
Consider the steering scenario where Alice steers Bob. Suppose that they want to reproduce exactly an arbitrary assemblage $\sigma$. Then, the length $t$ of the message is lower bounded by
\begin{align}
t\geq \log_2[\nu(\sigma)+1] \, .
\end{align}
Here $\nu(\sigma)$ is a quantifier of steering termed the LHS robustness, defined as the minimum $\mu\in\mathbb{R}_{\geq 0}$ such that the assemblage
\begin{equation}
\label{eq:defrobust}
\sigma^*=\frac{1}{1+\mu} \sigma+\frac{\mu}{1+\mu}\, \tilde{\sigma}
\end{equation}
is an LHS assemblage, with $\tilde{\sigma}$ any LHS assemblage.
\end{result}
We note that the LHS robustness $\nu$ is a variant of a quantifier introduced in \cite{piani}, and it is a steering monotone \cite{rodrigo14}.
\begin{proof}[Proof of Result \ref{theo3}]
Let us suppose that $\sigma$ can be simulated with a protocol using a message of length $t$. The strategy of the proof is to show that this implies that the assemblage
\begin{align}
\label{eq:unasse}
\sigma^*= \frac{1}{2^t} \, \sigma + \left( 1 - \frac{1}{2^t} \right) \,\tilde{\sigma}  \,.
\end{align}
is unsteerable for an unsteerable $\tilde{\sigma}$. Rewriting Eq. \eqref{eq:unasse} as 
\begin{align}
\sigma^*= \frac{1}{1+(2^t-1)} \, \sigma + \frac{(2^t-1)}{1+(2^t-1)} \,\tilde{\sigma} \,,
\end{align}
identifying $\mu \rightarrow 2^t-1$ implies $\nu(\sigma_{a|x})  \leq 2^t -1$, and the claim follows. 

Hence, it only remains to be shown that indeed $\sigma^*$ is unsteerable. For this, we show that the communication protocol to simulate $\sigma$ with a message of length $t$ can be recast into a LHS model of the form \eqref{eq:lhs} for $\sigma^*$. The details are provided in App. \ref{App_A}, but the main idea to construct the LHS model is the following. Suppose $t$ is the length of the message that Alice has to send Bob to simulate $\sigma$. Using the shared randomness $\lambda$, they now choose one $\mathbf{m}$ among all the possible length $t$ messages. Bob then prepares the state $\sigma_{\mathbf{m},\lambda}$. Alice, in turn, is given an input $x$ and should provide an output $a$: the way she produces this outcome is the following. If the message $\mathbf{m}$ is the one she would send Bob for this $x$ when simulating $\sigma$, she outputs $a$ according to the simulation protocol of the target assemblage. This happens with probability $\frac{1}{2^t}$ and produces  $\sigma$. Otherwise, she outputs $a$ according to an arbitrary but fixed $\tilde{p}(a|x)$. This happens with probability $1 - \frac{1}{2^t}$ and produces an unsteerable assemblage $\tilde{\sigma}$. Altogether, this reproduces $\sigma^*$, proving that it is indeed unsteerable, which concludes the proof.
\end{proof}

To illustrate the relevance of the method, we obtain lower bounds on the communication cost $t$ for a class of full-rank mixed entangled states, namely two-qudit isotropic states: $\varrho(V) = V \ket{\phi^+_d} \bra{\phi^+_d} + (1-V) \openone / d^2$, where $\ket{\phi^+_d} = \sum_{i=0}^{d-1}\ket{i}\ket{i}$ and $V$ is the visibility. To do so, we use the fact that $\nu(\sigma)$ can alternatively be shown to be equal to the largest possible normalised violation of any linear steering inequality. We then use the recent results on unbounded violations of steering inequalities \cite{HorodeckiUnbounded} to find strong lower bounds. In particular, in prime-power dimension, by measuring $(d+1)$ mutually-unbiased bases, we find $t \geq \log_2(V\sqrt{d}/2)$. Similarly, by measuring $\log_2(d)$ dichotomic Clifford observables, we find $t \geq \log_2(V\sqrt{\log_2(d)/2}/2)$. In both cases, the amount of communication is seen to grow with the dimension $d$, even though a finite number of measurements are made, and is robust in the presence of white noise. All details can be found in App. \ref{App_B}.

\emph{Approximate simulation.}--- Finally, we consider the question of how much communication is needed for the approximate simulation of an assemblage. More precisely, given an approximation error $\epsilon > 0$, what is the minimum length $t$ of the message from Alice to Bob such that each element $\sigma^{(\mathrm{sim})}_{a|x}$ of Bob's assemblage, given by Eq.~\eqref{assem cc}, is close in trace distance to the target one $\sigma_{a|x}$. That is, we demand that,
\begin{align}
\label{trace_dist}
\mathrm{d}\left(\varrho_{a|x}, \varrho^{(\mathrm{sim})}_{a|x}\right) \leq\epsilon, \, \forall\, a,x \text{ s.t. } p(a|x)>0,
\end{align}
where $\mathrm{d}(\varrho, \varrho')=\frac{1}{2}\|\varrho - \varrho'\|$ denotes the trace distance between $\varrho$ and $\varrho'$, and $\varrho^{(\mathrm{sim})}_{a|x}=\sigma^{(\mathrm{sim})}_{a|x}/\tr[\sigma^{(\mathrm{sim})}_{a|x}]$ is the normalised simulated conditional state. 

An upper bound on $t$ can be readily obtained from so-called epsilon nets \cite{Hayden04}, i.e. sets of pure states for which Eq.~\eqref{trace_dist} is known to hold. Here, however, we are interested in lower-bounding $t$, which must necessarily take into account the general situation where the conditional states $\varrho^{(\mathrm{sim})}_{a|x}$ can be mixed. This is the content of our next result.

\begin{result}
\label{Result4}
Consider the steering scenario where Alice steers Bob by performing projective rank-1 measurements. Suppose that they want to reproduce approximately the assemblage that arises from all such measurements on a pure-entangled state of two qubits $\ket{\psi_\theta}$.
Then, 
\begin{align}
\label{t_min}
t>t_\mathrm{bound} = \log_\mathrm{2} \left(\frac{2}{1 - \sqrt{1-4\epsilon^2}}\right)\,,
\end{align}
where $\epsilon > 0$ is the tolerated error.
\end{result}
We note that, for small values of $\epsilon$, a Laurent expansion of the right-hand side of Eq.~\eqref{t_min} yields 
\begin{align*}
t_\mathrm{bound} \approx \log_\mathrm{2} \left(\frac{1}{\epsilon^2}\right)\,.
\end{align*}
Clearly, $t_\mathrm{bound}\to\infty$ for $\epsilon\to0$, thus recovering Result \ref{theo1}. The proof of Result \ref{Result4} is given in App. \ref{App_C}.


\emph{Conclusion.}--- We discussed the communication cost of quantum steering, i.e. the minimal amount of classical communication required to simulate a quantum steering experiment without using any entanglement. In particular, we demonstrated that this communication cost is infinite for any pure bipartite entangled state. This further confirms that steering can be considered a form of quantum nonseparability which is intermediate between entanglement and Bell nonlocality, revealing striking differences between these concepts.

Also, we showed that the communication cost of steering is infinite even for certain mixed (non full-rank) entangled states. While this cannot be the case for all non full-rank entangled states, as some of these admit a LHS model \cite{finite1,finite2}, it would be still be interesting to see if there exist full-rank entangled states with infinite communication cost.  

Moreover, we showed how the communication cost of steering can be lower bounded in general, for arbitrary mixed-state assemblages and also in approximate simulations. In the future it would be interesting to find methods for placing upper bounds on the communication cost. In particular, one could consider the problem of constructing explicit LHS models assisted with $t$ bits of classical communication in order to simulate steering for given entangled mixed states. Notably, such a model was presented very recently for the case $t = 1$ and two-qubit Werner states \cite{tamas16}.

Finally, one may study the communication cost in multipartite steering experiments \cite{he,Cav15}, where many different communication patterns can be considered. In particular, it would be interesting to investigate the effect of post-quantum steering \cite{postQ}, which is only possible in the multipartite case.


\emph{Note added.}--- While finishing this work, we became aware of related and complementary work by Nagy and V\'ertesi \cite{tamas16}.

\emph{Acknowledgements.}--- We thank T. V\'ertesi for discussions, and acknowledge financial support from: the Swiss National Science Foundation (grant PP00P2\_138917 and Starting Grant DIAQ); the Beatriu de Pin\'os fellowship (BP-DGR 2013); ERC AdG NLST; the Brazilian agencies CNPq, FAPERJ, and INCT-IQ; and the Alexander von Humboldt foundation.


\appendix
\section{Details of the proof of Result \ref{theo2}}
\label{App_A}

In this appendix we present the details of the classical protocol without communication that Alice and Bob use in the ``Minimum communication cost for arbitrary assemblages'' section to construct the assemblage $\sigma^*$. This assemblage is a probabilistic mixture of the ``target'' assemblage $\sigma$, with probability $\frac{1}{2^t}$, and an unsteerable one $\tilde{\sigma}$, with probability $1 - \frac{1}{2^t}$, where $t$ is the length of the message that Alice should send Bob to simulate $\sigma$.

First of all, since the target assemblage $\sigma$ can be simulated by Alice and Bob when she is allowed to send him a message $\mathbf{m}$ of $t$ bits, it holds that $\sigma=\sigma^{(\mathrm{sim})}$, with $\sigma^{(\mathrm{sim})}$ given in Eq.~\eqref{assem cc}. Without loss of generality we can assume that the choice of message ${\bf m}$ is deterministic on $\hat{\mathbf{x}}$ and $\lambda$, hence: 
\begin{align}
\label{eq:asseccd}
\sigma_{a|\hat{\mathbf{x}}} = \int d\lambda \, \mu(\lambda) \, p(a|\hat{\mathbf{x}},\lambda) \, \varrho_{{\bf m}, \lambda}\,.
\end{align}

Now, the protocol to produce $\sigma^*$ goes as follows:
\begin{itemize}
\item[-] Alice and Bob choose uniformly at random a message $\boldsymbol{\tilde{m}}$ from a set of $\{0,1\}^t$ prefixed messages, aided by the shared randomness $\lambda$. 
\item[-] Bob: prepares the normalised state $\varrho_{\boldsymbol{\tilde{m}}, \lambda}$. 
\item[-] Alice: given $x$ and $\lambda$, if $\boldsymbol{\tilde{m}}= {\bf m}$ (i.e., the message she'd send Bob given $x$ and $\lambda$ when simulating $\sigma$) outputs $a$ with the probability $p(a|\hat{\mathbf{x}},\lambda)$ given in Eq. ~(\ref{eq:asseccd}). Otherwise, she outputs $a$ according to an arbitrary but fixed probability $\tilde{p}(a|\hat{\mathbf{x}})$. 
\end{itemize}

The unsteerable assemblage produced by this protocol has the following form: 
\begin{align}
\label{eq:sigma_star}
\nonumber
\sigma^*_{a|\hat{\mathbf{x}}} &= \frac{1}{2^t} \int d\lambda\,  \mu(\lambda) \, p(a|\hat{\mathbf{x}},\lambda) \, \varrho_{{\bf m}, \lambda} \\
&+ \frac{1}{2^t} \int d\lambda\,  \mu(\lambda) \sum_{\boldsymbol{\tilde{m}} \neq {\bf m}} \tilde{p}(a|\hat{\mathbf{x}}) \, \varrho_{\boldsymbol{\tilde{m}},\lambda} \,.
\end{align}
The first term clearly prepares $\sigma$ with probability $\frac{1}{2^t}$. Now we prove that the second term prepares an unsteerable assemblage $\tilde{\sigma}$, with components 
\begin{align}
\label{eq:tilde_sigma}
\tilde{\sigma}_{a|\hat{\mathbf{x}}}=\frac{1}{2^t-1}\int d\lambda \, \mu(\lambda) \sum_{\boldsymbol{\tilde{m}} \neq {\bf m}} \tilde{p}(a|\hat{\mathbf{x}}) \, \varrho_{\boldsymbol{\tilde{m}},\lambda},
\end{align}
with probability $1 - \frac{1}{2^t}$. 

Because of the above preparation protocol, $\tilde{\sigma}$ is guaranteed to be LHS, we just need to prove that it is a well-defined assemblage. First, we need to prove that $\sum_a\tilde{\sigma}_{a|\hat{\mathbf{x}}}=\frac{1}{2^t-1}\int d\lambda \, \mu(\lambda) \sum_{\boldsymbol{\tilde{m}} \neq {\bf m}} \varrho_{\boldsymbol{\tilde{m}},\lambda}$ is positive semidefinite and independent of $\hat{\mathbf{x}}$, despite the fact that $ {\bf m}$ does depend on $\hat{\mathbf{x}}$. The former is clear, since the $\varrho_{\boldsymbol{\tilde{m}},\lambda}$ are positive semidefinite themselves. For the latter, notice that
\begin{align}
\sigma_{a|\hat{\mathbf{x}}} = \int d\lambda \, \mu(\lambda) \, p(a|\hat{\mathbf{x}},\lambda) \, \left( \sum_{\boldsymbol{\tilde{m}}} - \sum_{\boldsymbol{\tilde{m}} \neq {\bf m}} \right) \varrho_{\boldsymbol{\tilde{m}}, \lambda}\,,
\end{align}
which implies
\begin{align}
\label{eq:indep}
\int d\lambda\,  \mu(\lambda) \sum_{\boldsymbol{\tilde{m}} \neq {\bf m}} \varrho_{\boldsymbol{\tilde{m}},\lambda} = \int d\lambda  \mu(\lambda) \,  \sum_{\boldsymbol{\tilde{m}}} \varrho_{\boldsymbol{\tilde{m}}, \lambda} - \varrho_\mathrm{B} \,,
\end{align}
where $\varrho_{\mathrm{B}}=\sum_a \sigma_{a|\hat{\mathbf{x}}}$ is the reduced state on Bob's lab. Since the right hand size of Eq.~(\ref{eq:indep}) is independent of $\hat{\mathbf{x}}$, we prove our first claim. 
Second, we need to prove that $\tilde{\sigma}$ is well normalised. We do this by noticing that the left hand size of Eq.~(\ref{eq:indep}) has trace $2^t-1$. 

From this, and using Eqs. \eqref{eq:asseccd}, \eqref{eq:sigma_star} and \eqref{eq:tilde_sigma}, Eq. \eqref{eq:unasse} finally follows, which finishes the proof.
\section{Communication cost for isotropic states}
\label{App_B}

In this appendix we present the details of the lower bound on the communication cost for the example full-rank states given in the main text. First, we elaborate on an equivalent definition of the LHS robustness and then compute the lower bound on the communication cost for simulating two particular assemblages. 

The LHS robustnes of steering $\nu$ defined by \eqref{eq:defrobust} can be expressed as a semidefinite program (SDP) in the following way:
\begin{align*}
1 + 2 \, \nu(\sigma) = &\min_{\{\varrho_\lambda, \tilde{\rho}_\lambda\}} \, \tr\left({\sum_\lambda \varrho_\lambda}\right) + \tr\left({\sum_\lambda \tilde{\rho}_\lambda}\right) \\
 &\mathrm{s.t.} \quad \sigma_{a|x} + \sum_\lambda D(a|x,\lambda)\, \varrho_\lambda \\
 &\qquad\qquad=  \sum_\lambda D(a|x,\lambda)\, \tilde{\rho}_\lambda \quad \forall \, a,x \,, \\ 
& \quad \quad \varrho_\lambda \geq 0\,, \quad \tilde{\rho}_\lambda \geq 0 \quad \forall \, \lambda \,,
\end{align*}
where $D(a|x,\lambda)$ are deterministic conditional probability distributions. 

When moving on to the dual of such an SDP, the LHS robustness can be computed via
\begin{align}
1 + 2 \, \nu(\sigma) \quad = \quad &\max_{\{F_{a,x}\}} \, \tr\left({\sum_{a,x} F_{a,x}  \,\sigma_{a|x}}\right) \\
 &\mathrm{s.t.} \quad \mathbbm{1} \geq \sum_{a,x} F_{a,x} \, D(a|x,\lambda) \geq -\mathbbm{1} \quad \forall \, \lambda\,. \label{dualC}
\end{align}
Hence the LHS robustness can be obtained by maximising the violation of a steering inequality provided that its LHS bound satisfies $|\beta_\mathrm{LHS}| \leq 1$. 

The following examples rely on steering inequalities with unbounded quantum violations. They both focus on the steering scenario where Alice steers Bob by performing $m$ measurements on the $d$-dimensional isotropic state
\begin{equation*}
\varrho_V = V \, \ket{\phi^+_d}\bra{\phi^+_d} + (1-V) \, \frac{\mathbbm{1}}{d^2}\,,
\end{equation*}
where $\ket{\phi^+_d}$ is the maximally entangled state in dimension $d$.

\quad 

{\it Projective measurements via MUBs.--}  Consider the steering scenario where Alice steers Bob by performing $m$ projective $d$-outcome measurements $\Pi_{a|x}$ on the isotropic state of dimension $d$. Assume moreover that these measurements are given by the $m=d+1$ mutually unbiased basis in $\mathbb{C}^d$.

The assemblage that is produced has the form
\begin{equation}
\sigma_{a|x} = \frac{V}{d} \, \Pi^T_{a|x} + \frac{1-V}{d} \, \frac{\mathbbm{1}}{d} \,.
\end{equation}

Now consider the particular steering inequality given by the operators $F_{a,x} = \frac{\Pi_{a|x}}{1+(m-1)/\sqrt{d}}$. The result by \cite{HorodeckiUnbounded} implies that these $F_{a,x}$ satisfy constraint (\ref{dualC}), hence the violation of the corresponding steering inequality by $\sigma_{a|x}$ will lower bound its LHS robustness. More precisely, 
\begin{align*}
\nonumber
1 + 2 \, \nu(\sigma) &\geq& \tr\left({\sum_{a,x} F_{a,x}  \,\sigma_{a|x}}\right) \\
&=& \frac{1+d}{1+\sqrt{d}} \left[ \frac{1}{d} + V \, \left( 1 - \frac{1}{d} \right) \right] \,.
\end{align*}

Hence a lower bound on the communication required to simulate this $\sigma_{a|x}$ is
\begin{align*}
t \geq \log_2 \left\{ \frac{1+d}{1+\sqrt{d}} \left[ \frac{1}{d} + V \, \left( 1 - \frac{1}{d} \right) \right] + 1 \right\} - 1\,. 
\end{align*}

In the limit of large dimensions, that is, a large number of measurements and outcomes, this bound behaves as
\begin{align*}
t \geq \log_2 \left( \frac{V \sqrt{d}}{2} \right)\,.
\end{align*}

\quad

{\it Clifford operators assemblages.--}  Consider the steering scenario where Alice steers Bob by performing $m=\log_2(d)$ dichotomic measurements $M_{a|x} = \frac{\mathbbm{1} + (-1)^{a+1} A_x}{2}$ on the isotropic state of dimension $d$. Assume that these operators $A_x$ are traceless Hermitian operators $\{A_i\}_{i=1:m}$ on the Hilbert space $\mathcal{H}_\mathrm{d}$, with the property that $A_x^2 = \, \mathbbm{1}_\mathrm{d}$ and that they anticommute. Such a set is given by Hermitian operators among the generators of a Clifford Algebra \cite{HorodeckiUnbounded}.

The assemblage produced by Alice has the form
\begin{equation}
\sigma_{a|x} = \frac{V}{d} \, \frac{\mathbbm{1} + (-1)^{a+1} A_x}{2} + \frac{1-V}{d} \, \frac{\mathbbm{1}}{2} \,.
\end{equation}

Now consider the particular steering inequality given by $F_{a,x} = \frac{(-1)^{a+1}}{\sqrt{2m}} \, A_x$.  The result by \cite{HorodeckiUnbounded} implies that these $F_{a,x}$ satisfy constraint (\ref{dualC}), hence the violation of the corresponding steering inequality by $\sigma_{a|x}$ will lower bound its LHS robustness. That is,
\begin{align*}
1 + 2 \, \nu(\sigma) \geq \tr\left({\sum_{ax} F_{a,x}  \,\sigma_{a|x}}\right) = V \sqrt{\frac{m}{2}} \,.
\end{align*}

A lower bound on the communication required to simulate this $\sigma_{a|x}$ is hence
\begin{align*}
t \geq \log_2 \left( \frac{V \sqrt{m/2}+1}{2} \right) \,, 
\end{align*}
which in the limit of large dimensions, that is, a large number of dichotomic measurements, behaves as
\begin{align*}
t \geq \log_2 \left( \frac{V \sqrt{\log_2(d)/2}}{2} \right)\,.
\end{align*}

\section{Proof of Result \ref{Result4}}
\label{App_C}

We recall here the scenario laid out in Result \ref{theo1}, where the measurement $x$ is identified with a direction in the Bloch sphere $\hat{\mathbf{x}}$ of the corresponding projector of Alice's measurement on the maximally entangled state. The assemblage that has to be reproduced approximately, $\sigma_{a|\hat{\mathbf{x}}}$ is given by Eq. \eqref{eq:rank1}. In this case, Eq. \eqref{trace_dist} takes the form
\begin{align}
\label{trace_distap}
\mathrm{d}\left(\ketbra{\Phi_{a|\hat{\mathbf{x}}}}{\Phi_{a|\hat{\mathbf{x}}}}, \varrho^{(\mathrm{sim})}_{a|\hat{\mathbf{x}}}\right) \leq\epsilon, \, \forall\, a,\hat{\mathbf{x}} \text{ s.t. } p(a|\hat{\mathbf{x}})>0.
\end{align}

The proof strategy consists in showing that if Eq. \eqref{t_min} is not fulfilled, the maximal number $N_\mathrm{max}$ of different possible states $\varrho^{(\mathrm{sim})}_{a|\hat{\mathbf{x}}}$ in the assemblage $\sigma^{(\mathrm{sim})}=\big\{p(a|\hat{\mathbf{x}})\,\varrho^{(\mathrm{sim})}_{a|\hat{\mathbf{x}}}\big\}$ with which Alice and Bob try to cheat is not enough for Eq. ~\eqref{trace_distap} to hold.

First, note that each $\varrho^{(\mathrm{sim})}_{a|\hat{\mathbf{x}}}$ can be written in Bloch representation as $\varrho^{(\mathrm{sim})}_{a|\hat{\mathbf{x}}}=\frac{\mathbbm{1} + \boldsymbol{\gamma} \cdot\boldsymbol{\sigma}}{2}$, where $\boldsymbol{\gamma}=\boldsymbol{\gamma}(a,\hat{\mathbf{x}})\in\mathbb{R}^3$ is the Bloch vector of $\varrho^{(\mathrm{sim})}_{a|\hat{\mathbf{x}}}$, with Euclidean norm $\gamma\leq1$, and $\boldsymbol{\sigma}$ is the Pauli operator vector. Using Eq. \eqref{trace_distap} and the fact that the Bloch vector of $\varrho_{a|\hat{\mathbf{x}}}=\ketbra{\Phi_{a|\hat{\mathbf{x}}}}{\Phi_{a|\hat{\mathbf{x}}}}$ has unit Euclidean norm (because it is pure), we get $\mathrm{d}\left(\varrho_{a|\hat{\mathbf{x}}}, \varrho^{(\mathrm{sim})}_{a|\hat{\mathbf{x}}}\right)=\frac{\sqrt{1-2\gamma \cos \theta + \gamma^2}}{2}$, where $\theta$ is the angle between the two Bloch vectors in question. Due to Eq.~\eqref{trace_distap}, $\varrho^{(\mathrm{sim})}_{a|\hat{\mathbf{x}}}$ will yield a valid approximate simulation of a given $\varrho_{a|\hat{\mathbf{x}}}$ only if $\theta$ is such that
\begin{equation}
\label{constraint}
\frac{\sqrt{1-2\gamma \cos \theta + \gamma^2}}{2} \leq \epsilon.
\end{equation}

We want each $\varrho^{(\mathrm{sim})}_{a|\hat{\mathbf{x}}}$ to approximately simulate as large an area on the surface of the Bloch sphere as possible. That is, we wish to find the optimal length $\gamma_\mathrm{min}$ of $\boldsymbol{\gamma}$ that maximises $\theta$. To this end, we minimise $\cos \theta_\mathrm{max}$ over $\gamma$ subject to the constraint \eqref{constraint}. This gives
\begin{equation}
\label{cos_max}
\cos \theta_\mathrm{max} = \sqrt{1 - 4\epsilon^2},
\end{equation}
which is attained at the optimal length $\gamma_\mathrm{min} = \sqrt{1-4\epsilon^2}$. 
The latter maximal angle corresponds to a total solid angle $2\pi\,(1-\cos \theta_\mathrm{max})$, centred at the direction of $\boldsymbol{\gamma}$, on the surface of the Bloch sphere. 

Then, clearly, if 
\begin{equation}
N_\mathrm{max}<\frac{4\pi}{2\pi\,(1-\cos \theta_\mathrm{max})}=2^{t_\mathrm{bound}}, 
\end{equation}
the simulation will not satisfy Eq. \eqref{trace_distap}. Furthermore, even $N_\mathrm{max}=2^{t_\mathrm{bound}}$ is not sufficient either. To see this, notice that the latter case corresponds to the well-known geometrical problem of packing equally-sized circles on the surface of a sphere \cite{Well91}. In particular, packing the surface of the unit sphere -- of total area $4\pi$ -- with  $2^{t_\mathrm{bound}}$ circles of area $4\pi/2^{t_\mathrm{bound}}$ cannot be done without the circles overlapping and, therefore, necessarily leaving uncovered areas. This, in turn, implies that there are pure target states $\varrho_{a|\hat{\mathbf{x}}}$ from which the closest $\varrho^{(\mathrm{sim})}_{a|\hat{\mathbf{x}}}$ will be further away than $\epsilon$. Hence, it must hold that 
\begin{equation}
\label{Nmax}
N_\mathrm{max}>2^{t_\mathrm{bound}}. 
\end{equation}

Now, as in the proof of Theorem \ref{theo1}, since $\lambda$ and $\hat{\mathbf{x}}$ are independent, the only way to have a successful simulation with $N_\mathrm{max}$ different states $\varrho^{(\mathrm{sim})}_{a|\hat{\mathbf{x}}}$ is that, for at least one value $\lambda^{*}$, there are already $N_\mathrm{max}$ different states $\varrho^{(\mathrm{sim})}_{a|\hat{\mathbf{x}},\lambda^*}$. Conditioned on an arbitrary value $\lambda^*$, the simulated average state is given by the mixture $\varrho^{(\mathrm{sim})}_{a|\hat{\mathbf{x}},\lambda^*}=\sum_{\bf m} q({\bf m}|\hat{\mathbf{x}},\lambda^*)\,  \varrho_{{\bf m},\lambda^*}$, but, also as in the proof of Theorem \ref{theo1}, there are at most $2^t$ distinct such mixtures. Hence, it must hold that $2^t\geq N_\mathrm{max}$, which, together with Eq.~\eqref{Nmax}, renders Eq.~\eqref{t_min} true.

\end{document}